\documentclass{eptcs}

\usepackage[latin1]{inputenc}
\usepackage{enumerate}
\usepackage{verbatimbox}
\usepackage{alltt}
\usepackage{eepic}
\usepackage{epic}
\usepackage{latexsym}
\usepackage{amssymb,stmaryrd}
\usepackage{amsmath}  
\usepackage{color,epsfig}
\usepackage{wrapfig}
\usepackage{float}
\usepackage{verbatim}
\usepackage{alg}
\usepackage{msc}
\usepackage{verbatim}
\usepackage{subfigure}
\usepackage{latexsym}

\title{Synthesis of Parametric Programs using Genetic Programming and Model Checking}

\author{Gal Katz \qquad\qquad Doron Peled%\\(\longrightarrowday)}
\institute{Department of Computer Science, Bar Ilan University\\
Ramat Gan 52900, Israel}
}
\def\titlerunning{Synthesis of Parametric Programs using Genetic Programming and Model Checking}
\def\authorrunning{G. Katz \& D. Peled}

\begin{document}
\maketitle

\begin{abstract}
Formal methods apply algorithms based on mathematical principles to
enhance the reliability of systems.
It would only be natural to try to progress from
verification, model checking or testing a system against its formal 
specification into constructing it automatically.
Classical algorithmic synthesis theory provides interesting algorithms
but also alarming high complexity and undecidability results.
The use of genetic programming, in combination of
model checking and testing, provides a powerful heuristic to synthesize
programs. The method is not completely automatic, as it is fine tuned
by a user that sets up the specification and parameters. It
also does not guarantee to always succeed and converge towards a
solution that satisfies all the required properties. However,
we applied it successfully on quite nontrivial examples
and managed to find solutions to hard programming challenges, as well
as to improve and to correct code.
We describe here several versions of
our method for synthesizing sequential and concurrent systems. 
\end{abstract}

\section{Introduction}

Formal methods~\cite{SRM} assist 
software and hardware developers in enhancing the reliability of systems.
They provide methods and tools to search for design and programming errors.
While these methods are effective in the software development
process, they also suffer from severe limitations: testing is not
exhaustive, formal verification is extremely tedious and 
model checking is limited to particular domains (usually, finite
state systems) and suffers from high complexity,
where memory and time requirements are sometimes
prohibitively high.

A natural progress from formal methods 
are algorithms for automatically converting
the formal specification into code or a description of hardware.
Such algorithms would create correct-by-design code or piece of hardware. 
However, high complexity~\cite{PR1} and even undecidability~\cite{PR2} appear
in some main classical automatic synthesis problems.

The approach presented here is quite different from
algorithmic synthesis. We perform a generate-and-check
kind of synthesis and use model checking or
SAT solving to evaluate the generated candidates. 
An extreme approach would be to enumerate the possible programs
(say, up to a certain size) and 
use model checking to find the correct solution(s). This was applied in
Taubenfeld~\cite{BT} to find mutual exclusion algorithms.
Our synthesis method is based on {\em genetic programming}. It
allows us to generate multiple candidate solutions at random and to
mutate them, as a stochastic process. We employ enhanced model
checking (model checking that does not only produce an affirmation to
the checked properties or a counterexample, but distinguishes also some
finer level of correctness) to provide {\em fitness} levels that are used to
direct the search towards solutions that satisfy the given specification.
Our synthesis method
can be seen as a heuristic search in
the space of syntactically fitting programs. It is not completely automatic, in
the sense that the user can refine the specification and change the
way the fitness is evaluated when the formal properties are satisfied.
Our method is not guaranteed to terminate with a correct solution; we might
give up after some time and can restart the search from a new random seed 
or with a refinement of the way the method assigns fitness.

Although this marriage between genetic programming and model checking is quite
promising, it suffers from some limitations of model checking. First, 
model checking is primarily designed for finite state systems. 
Although some extensions of it exist (e.g., to programs with a single stack),
model checking does not work in general for parametric systems. Unfortunately,
most systems that we would like to synthesize are parametric in nature: 
almost every abstract algorithm on data structures, be it queue, tree, 
graph, is parametric,
where the size of the structure, is not fixed. 
It is easy to demonstrate model checking
on a sorting program with a fixed vector of numbers and 
some fixed initial assignment of
values. However, when the length of the vector is parametric, and we
need to prove correctness with respect to arbitrary 
set of values, existing model checking techniques often fail.

For this reason, we use model checking in our approach for
synthesizing parametric systems not as a comprehensive method
for finding correctness, but as a generalized testing tool, 
which can make exhaustive checks for fixed parameters.
Under this setting, 
we accept candidate programs when there is ample evidence that 
they are correct, specifically, when  they passed enough
checks, rather than when we establish comprehensive correctness.

Our genetic programming synthesis approach allows us not only to generate code
that satisfies a given temporal specification but also to improve and correct code.
We can start with an existing solution for a specification, and use the genetic
process to improve it. We can also start with some flawed version of the code and
use our method to correct it.

\section{Genetic Programming Based on Model Checking}

\begin{figure}[!htb]
	\centering
h		\includegraphics[scale=0.83]{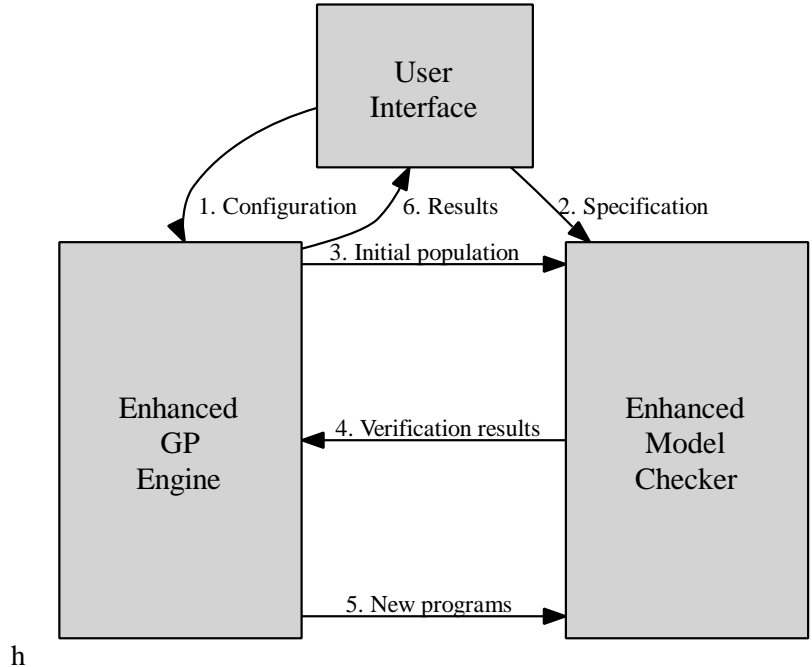}
	\caption{The Suggested Framework}
	\label{arch}
\end{figure}

We present in~\cite{KP1,KP2,KP4,KP5,KP6} a framework combining genetic 
programming and model checking, which allows to 
automatically synthesize code for given problems.
The framework we suggest is depicted in Figure~\ref{arch}.
\begin{itemize}
\item The {\em formal specification} of the problem, as 
well as the required architecture and constraints on the structure of 
the desired solutions is provided by the user. 
This may also include some initial versions of
the desired code that either need correction or improvement.
\item An \emph{enhanced GP engine} that generates random programs and then evolves them
using mutation operations that allow to change the code randomly.
\item A \emph{verifier} that analyzes the generated programs, 
and provides useful information 
about their correctness. This can be a model checker, often enhanced to
provide more information than yes/no (and counterexample), or a SAT solver.
\end{itemize}
The synthesis process goes through the following steps:
\begin{enumerate}
\item The user feeds the GP engine with the desired architecture and
a set of constraints regarding the programs that are allowed
to be generated. This includes:
\begin{enumerate}
\item a set of functions literals and instructions used as building blocks for the generated programs,
\item the number of concurrent processes, the methods of 
communication between processes (in case of concurrent programs), 
\item limitations on the size and structure of the generated programs, and
the maximal number of permitted iterations.
\item The user may also provide some initial versions of the code that may be
either incorrect or suboptimal. The genetic process can exploit these versions
to evolve into better (correct or optimized) code.

\end{enumerate}
\item The user provides a formal specification for the problem. 
This consists, in our case, of a set of linear temporal logic properties, 
as well as additional quantitative requirements on the program behavior.
\item The GP engine randomly generates an {\em initial population} of
candidate programs 
based on the provided building blocks and constraints. 
\item The verifier analyzes the behavior of the generated candidates against 
the specification properties, 
and provides {\em fitness measures} based on the amount of satisfaction.
\item The GP engine creates new
programs by applying the genetic operations of {\em mutation}, which performs
small random changes to the code, 
and {\em crossover}, which glues together parts of
different candidate solutions.
Steps~4 and~5 are repeated until either a perfect program is found 
(fully satisfying the specification), or until the maximal number of 
iterations is reached.
\item The results are sent back to the user. This includes  programs
that satisfy all the specification properties, if one exists, 
or the best partially correct programs that was found, 
along with its verification results.
\end{enumerate}

For steps~4 and~5 above, we use the following selection method:

\begin{itemize}
\item Randomly select $\mu$ candidate programs.
\item Create $\lambda$ new candidates by applying mutation (and optionally 
      crossover) operations (as explained below) to the above $\mu$ candidates.
      We now have $\mu + \lambda$ candidates.
\item Calculate the fitness function for each of the new candidates
based on ``enhanced model checking''.
\item 
Based on the calculated fitness, choose new $\mu$ candidates from the 
set of $\mu + \lambda$ candidates.
Candidates with higher fitness values are selected
with a higher probability than others.
Replace the originally selected $\mu$ with the ones selected 
at this step.
\end{itemize}
\label{gp-back}

We represent programs as trees, where an instruction or an expression is 
represented by a single node, having its parameters as its offspring. Terminal 
nodes represent constants. Examples of the instructions we use are 
\emph{assignment, while} \emph{, if} and \emph{block}. 
The latter construct is a sequential composition of a pair of instructions.

\begin{figure}
	\centering
		\includegraphics[scale=0.9]{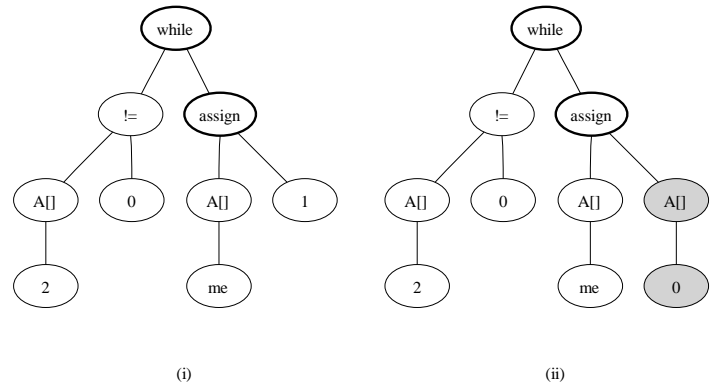}
	\caption{(i) Randomly created program tree, (ii) the result of a replacement mutation}
	\label{tree1rep}
\end{figure}

At the first step, an initial population of candidate programs is generated. 
Each program is generated recursively, starting from the root, adding 
nodes until the tree is completed. The root node is chosen randomly from the 
set of instruction nodes, and each child node is chosen randomly from the set 
of nodes allowed by its parent type, and its place in the parameter list.
Figure~\ref{tree1rep}(i) shows an example of a randomly created tree
that represents the following program:

\begin{minipage}{13cm}
\begin{verbatim}


    while (A[2] != 0)
        A[me] = 1

\end{verbatim}
\end{minipage}

\begin{figure}
	\centering
		\includegraphics[scale=0.5]{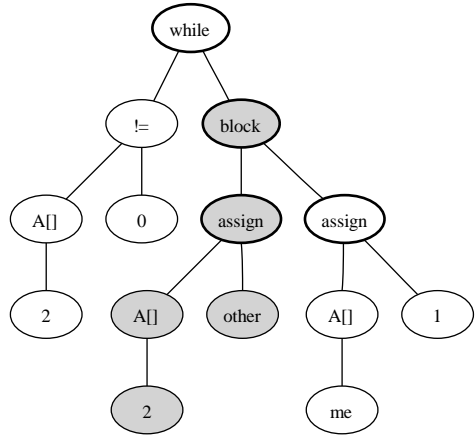}
	\caption{Tree after insertion mutation}
	\label{tree-insert}
\end{figure}

The main operation we use is {\em mutation}. It allows making small changes
on existing trees. The mutation includes the following steps:
\begin{enumerate}
\item Randomly choose a node $s$ from the program tree.
\item Apply one of the following operations to the tree with respect to 
ihe chosen node:
\begin{enumerate}
\item Replace the subtree with root $s$ with a new randomly generated subtree.
\item Add an immediate parent to $s$. 
Randomly create other offspring to the new parent, if needed.
\item Replace the node $s$ by one of its offspring. 
Delete the remaining offspring
of that node.
\item Delete the subtree with root $s$. The node ancestors should be 
updated recursively.
\end{enumerate}
\end{enumerate}
Mutation of type (a) can replace either a
single terminal or an entire subtree. For example, the terminal ``1'' in the
tree of Figure~\ref{tree1rep}(i), is replaced by the grayed subtree
in~\ref{tree1rep}(ii), 
changing the assignment instruction into \verb+A[me] = A[0]+.
Mutations of type (b) can extend programs in several ways, depending
on the new parent node type. In case a ``block'' type is chosen, a new
instruction(s) will be inserted before or after the mutation node.
For instance, the grayed part of Figure~\ref{tree-insert} represents a 
second assignment instruction inserted into the original program.
Similarly, choosing a parent node of type ``while'' will have the effect
of wrapping the mutation node with a while loop. 
The type of mutation applied to candidate programs
is randomly selected. All mutations must of course produce legal code.
This affects the possible mutation type for the chosen node, 
and the type of new generated nodes.

Another operation that is frequently used in genetic programming is {\em crossover}.
The crossover operation creates new candidates by merging building blocks of two existing programs. The crossover steps are:
\begin{enumerate}
\item Randomly choose a node from the first program.
\item Randomly choose a node from the second program that has the same type 
as the first node.
\item Exchange between the subtrees rooted by the two nodes, and use the two new programs created by this method.
\end{enumerate}
While traditional GP is heavily based on crossover, it is quite a
controversial operation (see \cite{gpintro}, for example). 
Crossover is not used in our work.

%We use another method, which is more similar to the 
%Evolutionary Strategies~\cite{es2} 
%$\mu+\lambda$ style. In this method, genetic operations are applied to all
%of the $\mu$ programs, in order to produce a much larger set of $\lambda$
%offspring. The fitness proportional selection is then used in order to chooses $\mu$ programs from the set of parents and 
%offspring that will replace the original $\mu$ parents.

{\em Fitness} is used by GP in order to choose which programs have a higher 
probability to survive and participate in the genetic operations. 
In addition, the success termination criterion of the GP algorithm is based 
on the fitness value of the most fitted candidate. Traditionally, 
the fitness function is calculated by running the program on some 
set of inputs (a training set), which represent the
possible inputs. In contrast, our fitness function is not 
based on running the programs on sample data, but on an enhanced model checking procedure.
While the classical model checking provides a yes/no answer to the 
satisfiablity of the specification (thus yielding a two-valued 
fitness function), our model checking algorithm generates a smoother
function by providing several levels of correctness.
%provides an alternative algorithm to
%the classical one, provinding more levels than
%just the standard {\em correct/incorrect}. 
Often, we have the following four levels of correctness, 
per each linear temporal logic property:
\begin{enumerate}
\item None of the executions of the program satisfy the property.
\item Some, but not all the executions of the program satisfy the property.
\item The only executions that do not satisfy the property must have
infinitely many decisions that avoid a path that does satisfy the property.
\item All the executions satisfy the property.
\end{enumerate}
We provided several methods for generating the various fitness levels: 
{\begin{itemize}
	\item Using Streett Automata, and a strongly-connected
	component analysis of the program graph~\cite{KP2}.
	\item Enhanced model checking logic and algorithm~\cite{KP1,NPP}.
	\item Probabilistic model checking.
\end{itemize}
%An algorithm for deep model checking, can be found in~\cite{NPP}. 

There are several other considerations in setting up the calculation of
the fitness. First, priority between the properties is used
to suppress assigning fitness value due to the satisfaction of a liveness 
property (e.g., ``when a process wants to enter its critical section, it would
eventually be able to do so'') when the safety property does not
hold (e.g., ``the two processes cannot enter their critical sections
simultaneously''). Another consideration is to prevent needless growth of
the program by useless code. To alleviate this, we use some negative
fitness value related to the program's length. This entails that a solution
that satisfies all the specification is accepted even if it does not
have perfect fitness value (due to length).

\section{Example: Mutual Exclusion Algorithms}

\label{example}
As an example, we used our method in order to automatically generate solutions to several variants of the Mutual Exclusion Problem. In this problem, first described and solved by Dijkstra \cite{Dijkstra}, two or more processes are repeatedly running critical and non-critical sections of a program. The goal is to avoid the simultaneous execution of the critical section by more than one process. We limit our search for solutions to the case of only two processes.
The problem is modeled using the following program parts that are executed in an infinite loop:

%\small
\begin{minipage}{13cm}
\begin{verbatim}


  Non Critical Section
  Pre Protocol
  Critical Section
  Post Protocol

\end{verbatim}
\end{minipage}
%\normalsize

These parts are fixed, and, together with the number of processes involved (two)
and the number of variables allowed, consist of the architecture provided to
our genetic programming tool, together with the temporal specification.

The \verb+Non Critical Section+ part represents the process part on 
which it does not require an access to the shared resource. A process can 
make a nondeterministic choice whether to stay in that part, or to move 
into the \verb+Pre Protocol+ part. 
From the \verb+Critical Section+ part, a process always has 
to move into the \verb+Post Protocol+ part. 
The \verb+Non Critical Section+ and \verb+Critical Section+ parts are fixed, 
while our goal is to automatically generate code for the  
\verb+Pre+ \verb+Protocol+ and \verb+Post Protocol+ parts, 
such that the entire program will fully satisfy the problem's specification.

We use a restricted high level language based on the C language.
Each process has access to its id (0 or 1) by the \verb+me+ literal, and to 
the other process' id by the \verb+other+ literal. The processes can use an 
array of shared bits with a size depended on the exact variant of the problem 
we wish to solve. The two processes run the same code. The available node 
types are: \emph{assignment, if, while, empty-while, 
block, and, or} and \emph{array}. Terminals include the constants: 
\emph{0, 1, 2, me} and \emph{other}.

Table~\ref{MutualSpecs} describes the properties that 
define the problem specification. 
The four program parts are denoted by \verb+NonCS+, \verb+Pre+, \verb+CS+ and \verb+Post+ respectively.
\begin{table}[ht]
	\centering
	\small
		\caption{Mutual Exclusion Specification}
		\begin{tabular}{|l|l|p{5.5cm}|l|l|}
		\hline
		No. & Type & Definition & Description & Level \\
		\hline \cline{1-5}
		1	& Safety & $\Box \lnot (p_{0}$ in \verb+CS+ $\land\ p_{1}$ in \verb+CS+) & 			Mutual Exclusion & 1 \\ \hline
		2,3 & Liveness & $\Box (p_{me}$ in \verb+Post+ $\to \Diamond (p_{me}$ in \verb+NonCS+)) & Progress & 2 \\ \cline{1-1} \cline{3-5}
		4,5 & & $\Box (p_{me}$ in \verb+Pre+ $\land\ \Box (p_{other}$ in \verb+NonCS+)) $\to \Diamond (p_{me}$ in \verb+CS+)) & No Contest & 3 \\ \cline{1-1} \cline{3-5}
		6 & & $\Box ((p_{0}$ in \verb+Pre+ $\land\ p_{1}$ in \verb+Pre+) $\to \Diamond (p_{0}$ in \verb+CS+ $\lor\ p_{1}$ in \verb+CS+)) & Deadlock Freedom & 4 \\ 			\cline{1-1} \cline{3-5}
		7,8 && $\Box (p_{me}$ in \verb+Pre+ $\to \Diamond (p_{me}$ in \verb+CS+)) & Starvation & 4 			\\ \hline
	\end{tabular}
	\label{MutualSpecs}
	\normalsize
\end{table}
Property 1 is the basic safety property requiring the mutual exclusion. Properties displayed
in pairs are symmetrically defined for the two processes. Properties~2 and~3 
guarantee that the processes are not hung in the \verb+Post Protocol+ part. 
Similar properties for the \verb+Critical Section+ are not needed, since it 
is a fixed part without an evolved code. Properties~4 and~5 require that 
a process can enter the critical section, if it is the only process trying 
to enter it. Property~6 requires that if both processes are trying to enter 
the critical section, at least one of them will eventually succeed. 
This property can be replaced by the stronger requirements~7 and~8 that 
guarantee that no process will starve.

There are several known solutions to the Mutual Exclusion problem, depending on the number of shared bits in use, the type of conditions allowed 
(simple / complex) and whether starvation-freedom is required. The variants of the problem we wish to solve are showed in Table \ref{MutualVariants}.

\begin{table}[ht]
	\centering
	\small
		\caption{Mutual Exclusion Variants}
		\begin{tabular}{|p{1.1cm}|p{1.2cm}|p{1.6cm}|p{2.8cm}|p{1.7cm}|p{2.9cm}|}
			\hline
			Variant No. & Number of bits & Conditions	& Requirement & Relevant properties & Known algorithm \\
			\hline \cline{1-6}
			1 & 2	& Simple & Deadlock Freedom & 1,2,3,4,5,6 & One bit protocol \cite{one-bit} 
			\\ \hline
			2 & 3	& Simple & Starvation Freedom & 1,2,3,4,5,7,8 & Dekker \cite{Dijkstra} 
			\\ \hline
			3 & 3	& Complex	& Starvation Freedom & 1,2,3,4,5,7,8 & Peterson \cite{PetersonF77} 
			\\ \hline
		\end{tabular}
	\label{MutualVariants}
	\normalsize
\end{table}

Three different configurations where used, in order to search for solutions to the variants described in Table \ref{MutualVariants}. Each run included the creation of 150
initial programs by the GP engine, and the iterative creation of new programs until a perfect solution was found, or until a maximum of 2000 iterations. At each iteration, 5 programs were randomly selected, bred,
and replaced using mutation. 
The values $\mu=5,\lambda=150$ where chosen. 

%The tests were
%performed on a 2.6 GHz Pentium Xeon Processor. For each configuration, multiple runs
%were performed. Some of the runs converged into perfect solutions, while others found
%only partial solutions.
%The results are summarized on Table \ref{res-table}.
%\begin{table}[ht]
	%\centering
		%\small
		%\caption{Test results}
		%\begin{tabular}{|p{1cm}|p{2.5cm}|p{2.8cm}|p{2.9cm}|}
			%\hline
			%Variant No. & Successful runs (\%) & Avg. run durtaion (sec) & Avg. no. of tested programs per run\\
			%\hline \cline{1-4}
			%1 & 40 & 128 & 156600 \\
			%\hline
			%2 & 6 & 397 & 282300 \\
			%\hline
			%3 & 7 & 363 & 271950 \\
			%\hline
		%\end{tabular}
	%\label{res-table}
	%\normalsize
%\end{table}

In addition to the temporal specification of mutual exclusion,
our configuration allows three shared bits. 
The famous Dekker's algorithm \cite{Dijkstra} uses 
two bits to announce that they want to enter the critical section,
and the third bit is used to set turns
between the two processes. 
Many runs initially converged into
deadlock-free algorithms using only two bits. Those algorithms have 
executions in which one of
the processes starve, hence only partially satisfying properties 7 or 8. 
Program (a) shows one of those algorithms, which later evolved into program (b). 
The evolution first included the addition of the second line 
to the \emph{post protocol} section (which 
only slightly decreased its fitness level due to the parsimony measure). A replacement
mutation then changed the inner while loop condition, leading to a 
perfect solution similar to Dekker's algorithm.

\begin{center}
%\footnotesize
\begin{minipage}{11.4cm}
\hspace{-3cm}
\begin{verbatim}

Non Critical Section       Non Critical Section    
A[me] = 1                  A[me] = 1              
While (A[other] == 1)      While (A[other] == 1)  
   While (A[0] != other)      While (A[2] == me)  
      A[me] = 0                  A[me] = 0        
   A[me] = 1               A[me] = 1             
Critical Section           Critical Section        
A[me] = 0                  A[2] = me
                           A[me] = 0

    (a) [94.34]              (b) [96.70]          
\end{verbatim}
\end{minipage}
%\normalsize
\end{center}

%The first problem for which we wanted to synthesize solutions was the
%classical {\em mutual exclusion} problem~\cite{Dijkstra}. 
%The temporal specification (in 
%Linear Temporal Logic) for the problem are given in Table~\ref{MutualSpecs}.
%
%\begin{table}[ht]
	%\centering
	%\small
		%\caption{Mutual Exclusion Specification}
		%\begin{tabular}{|l|l|p{5.5cm}|l|}
		%\hline
		%No. & Type & Definition & Description  \\
		%\hline \cline{1-4}
		%1	& Safety & $\Box \lnot (p_{0}$ in \verb+CS+ $\land\ p_{1}$ in \verb+CS+) & 			Mutual Exclusion  \\ \hline
		%2,3 & Liveness & $\Box (p_{me}$ in \verb+Post+ $\to \Diamond (p_{me}$ in \verb+NonCS+)) & Progress  \\ \cline{1-1} \cline{3-4}
		%4,5 & & $\Box (p_{me}$ in \verb+Pre+ $\land\ \Box (p_{other}$ in \verb+NonCS+)) $\to \Diamond (p_{me}$ in \verb+CS+)) & No Contest  \\ \cline{1-1} \cline{3-4}
		%6 & & $\Box ((p_{0}$ in \verb+Pre+ $\land\ p_{1}$ in \verb+Pre+) $\to \Diamond (p_{0}$ in \verb+CS+ $\lor\ p_{1}$ in \verb+CS+)) & Deadlock Freedom  \\ 			\cline{1-1} \cline{3-4}
		%7,8 && $\Box (p_{me}$ in \verb+Pre+ $\to \Diamond (p_{me}$
		%in \verb+CS+)) & Starvation Freedom 			\\ \hline
		%9 & Safety & $\Box\lnot$(\verb+remote writing+) & Single-Writer	\\ \hline
		%10 & Special & Bounded number of remote operations & Local-Spinning \\ \hline
	%\end{tabular}
	%\label{MutualSpecs}
	%\normalsize
%\end{table}

\vspace{1ex}
%Initially, we tried to rediscover three of the classical mutual exclusion algorithms:
%the \emph{one bit} protocol 
%and two starvation-free algorithms, \emph{Dekker's} and \emph{Peterson's} algorithms.
%Our framework (and tool) successfully discovered all of these 
%algorithms~\cite{KP1}, and even some interesting variants of them.

Inspired by algorithms developed by Tsay~\cite{Tsay} and by
Kessels~\cite{Kessels},
our next goal was to start from an existing algorithm, and by adding more constraints and building blocks, try
to evolve into more advanced algorithms.

First, we allowed a minor asymmetry between the two processes. 
This is done by the operators
{\em not0} and {\em not1}, which act only on one of the processes. Thus, for process $0$,
$\mathit{not0} ( x ) = \neg x$ while for process $1$,
$\mathit{not0} ( x ) = x$. This is reversed for
$\mathit{not1} ( x )$, which negates its bit operand $x$ only in process $1$, and do nothing on process $0$.

%For the first  algorithm, we used a similar configuration as in the
%our first paper~\cite{KP1}, i.e. 3 shared bits and
%allowing the constructs {\em if}, {\em while}, and {\em assignment}.
As a result, the tool found two algorithms that may be
considered simpler than Peterson's. The first one has only one condition in
the {\em wait} statement (written here using the syntax of
a {\em while} loop), although with a more complicated atomic comparison, 
between two bits.
Note that the variable {\sf turn} is in fact {\sf A[2]} and is
renamed here {\sf turn} to accord with classical presentation of
the extra global bit that does not belong to a specific process.

\begin{center}
%\footnotesize
\begin{minipage}{11.4cm}
\hspace{-3cm}
\begin{verbatim}
Pre CS
A[me] = 1
turn = me
While (A[other] != not1(turn));
Critical Section
A[me] = 0
\end{verbatim}
\end{minipage}
\end{center}
%\vspace{1ex}
The second algorithm uses the idea of setting 
the {\sf turn} bit one more time after
leaving the critical section. This allows the while condition to be even
simpler.  Tsay~\cite{Tsay} used a similar refinement, but 
his algorithm needs an additional {\sf if} statement, which is not used
in our algorithm.

\begin{center}
%\footnotesize
\begin{minipage}{11.4cm}
\hspace{-3cm}
\begin{verbatim}
Pre CS
A[me] = 1
turn = not0(A[other])
While (A[2] != me);
Critical Section
A[me] = 0
turn = other
\end{verbatim}
\end{minipage}
\end{center}
\vspace{1ex}
Next, we aimed at finding more advanced algorithms satisfying additional properties.
%added more constrains. 
The configuration was extended into four shared bits and two private bits (one for each process).
The first requirement was that each process can change only its 2 local
bits, but can read all of the 4 shared bits
This yielded the following algorithm.
\vspace{1ex}

\begin{center}
\begin{minipage}{11.4cm}
\hspace{-3cm}
\begin{verbatim}
Pre CS
A[me] = 1
B[me] = not1(B[other])
While (A[other] == 1 and B[0] == not1(B[1]));
Critical Section
A[me] = 0
\end{verbatim}
\end{minipage}
\end{center}
\vspace{1ex}
The algorithm uses the idea of using two bits as 
the ``turn'', were each process changes only its bit to set its turn, 
but compares both of them
in the while loop.
Finally, we added the requirement %to allow writing to all shared bits, but
for busy waiting only on local bits (i.e. using local spins). 
%It was implemented
%by a special kind of liveness property on the transitions without requiring
%fairness. 
The following algorithm (similar to Kessels') was generated, 
satisfying all properties from the table above.

\begin{center}
\begin{minipage}{11.4cm}
\hspace{-3cm}
\begin{verbatim}
Non Critical Section
A[other] = 1
B[other] = not1(B[0])
T[me] = not1(B[other])
While (A[me] == 1 and B[me] == T[me]);
Critical Section
A[other] = 0
\end{verbatim}
\end{minipage}
\end{center}
%Again, two bits are used for the ``turn'', but this time each process waits
%only on its bits.

\section{Synthesizing Parametric Programs}

Our experience with genetic program synthesis quickly hits a difficulty
that stems from the limited power of model checking: there are 
few interesting fixed finite state
programs that can also be completely specified using pure temporal logic.
Most programming problems are, in fact, parametric. Model checking
is undecidable for parametric families of programs
(say, with $n$ processes, each with the same code, initialized
with different parameters)
even for a fixed property~\cite{AK}. One may look at
mutual exclusion for a parametric number of processes. Examples are, sorting, 
where
the number of processes and the values to be sorted are the parameters,
network algorithms, such as finding the leader in a set of processes, etc.
In order to synthesize parametric concurrent programs, in particular
those that have a parametric number of processes, and even a parametric
architecture, we use a different genetic programming strategy.

First, we assume that a solution that is checked for a large number of
instances/parameters is acceptable. This is not a guarantee of
correctness, but under the prohibitive undecidability of model
checking for parametric programs, at least we have a strong evidence that 
the solution may generalize
to an arbitrary configuration. In fact, there are several works on
particular cases where 
one can calculate the parameter size that guarantees that
if all the smaller instances are correct, then any instance is
correct~\cite{EN}. Unfortunately, this is not a rule that can
be applied to any arbitrary parametric problem.
We apply a {\em co-evolution} based synthesis algorithm: 
we collect parameters from failed checked cases
and keep them as counterexamples. When suggesting a new solution,
we check it against the collected counterexamples.
We can view this process as a genetic search for both correct
programs and counterexamples. The fitness is different, of course,
for both tasks: a program gets higher fitness by being close to satisfying
the full set of properties, while a counterexample is obtaining
a high fitness if it fails the program.

In this sense, the
model checking of a particular set of instances can be considered as
a generalized {\em testing} for these values: 
each set of instances of the parameters provides a single finite
state system that is itself comprehensively tested using model checking. 
This idea can be also used, independently, for model checking parametric
systems.
For example, consider a concurrent sorting program 
consisting of a parametric array of processes, 
each containing some initial value. 
Adjacent processes may exchange values during the algorithm.
For any particular size and set of values, 
the model checking provides automatic and exhaustive test for a particular
set of values, but the check is not exhaustive for all the array sizes
or array values, but rather samples them.

In the classical {\em leader election
in a ring} problem, the processes initially have their own values 
that they can transfer around, with the goal of finding a process that 
has the highest
value. Then, the parameters include the size of the ring, and the initial
assignment of values to processes. While we can check solutions up to 
a certain size, and in addition, check all possible initial values,
the time and state explosion is huge. 
Instead, we can then store each set of instances of
the parameters that
failed for some candidate solution, and, when checking a new candidate 
solution,
check it against the failed instances. A solution for the leader election,
albeit not the most optimal one, was obtained using our genetic programming
methods~\cite{KP4}.

\section{Correcting Erroneous Program}
\label{sect:correcting}
Our method is not limited to finding new program that satisfy the
given specification. In fact, we can start with the code of an existing
program instead of a completely random population
and try to improve or correct it. In order to {\em improve code},
our fitness measure may include
some quantitative evaluation; then the initial program may be found inferior
to some later generated candidates. If the program we start with
is {\em erroneous}, then it would not get a very high fitness value by 
failing to satisfy some of the properties.

In~\cite{KP5} we approached the ambitious problem of correcting a known
protocol for obtaining interprocess interaction called
$\alpha$-core~\cite{alpha}. The algorithm allows multiparty
synchronization of several processes. It needs to function in a system
that allows nondeterministic choices, which makes it challenging, 
as processes that may consider one possible interaction may also
decide to be engaged in another interaction. The algorithm uses
asynchronous message passing in order to enforce selection of
the interactions by the involved processes. 
This nontrivial algorithm, which is used 
in practice for distributed systems, contains an error.

The protocol is quite big, involving sending different
messages between the controlled processes, and the controlling processes, 
one per each possible multiparty interaction. These messages include
announcing the willingness to be engaged in an interaction,
committing an interaction, canceling an interaction, request for
commit from the interaction manager processes, as well as announcement
that the interaction can start, or is canceled due to the
departure of at least one participant. 
The state space of such a protocol is obviously high.
In addition, the protocol can run on any number 
of processes, each process with arbitrary number of choices to
be involves in interactions, and each
interaction includes any number of processes.

Recall that model checking of parametric programs is undecidable 
in general~\cite{AK}. In fact, we use our genetic programming
approach first to find the error, and then to correct it.
We use two important ideas:
\begin{enumerate}
\item Use the genetic engine not only to generate programs, but also to evolve different architectures 
on which programs can run.
\item Apply a co-evolution process, where %architectures that
candidate programs, and test cases (architectures) that
may fail these programs, are evolved in parallel.
%show which suggested (and the original) candidate programs fail
%are used as test cases for further suggested candidates.
\end{enumerate}

Specifically, the architecture for the candidate programs is also
represented as code (or, equivalently, a syntactic tree) for
spanning processes and their interactions, which can be
subjected to genetic mutations. The fitness function directs the search for 
a program that may falsify
the specification for the given erroneous program.
After finding a ``bad'' architecture for a program, one that causes
the program to fail its specification, our next goal 
is to reverse the genetic programming direction, and try to automatically 
correct the program, where a ``correct'' program at this step, is one that has
passed model checking against the architecture. 
Yet, correcting the program for the first found
wrong architecture only, does not guarantee its correctness
under different architectures, hence more architectures that fail candidate
solutions are collected. Note that we use for the co-evolution
two separate fitness functions: one
for searching for ``bad'' architectures, and one for searching for
a correct solution.

In Figure~\ref{arch1} we show the architecture that was found to produce
the error in the original $\alpha$-core algorithm. A message sequence chart
in Figure~\ref{msc} demonstrate the found bad scenario. The correction
consisted of changing the line of code

\begin{center} {\sf if n$>$0 then n:=n-1} \end{center}

into
 
\begin{center} {\sf if sender$\in$shared then n:=n-1} \end{center}

\begin{figure}
	\centering
		\includegraphics[scale=0.65]{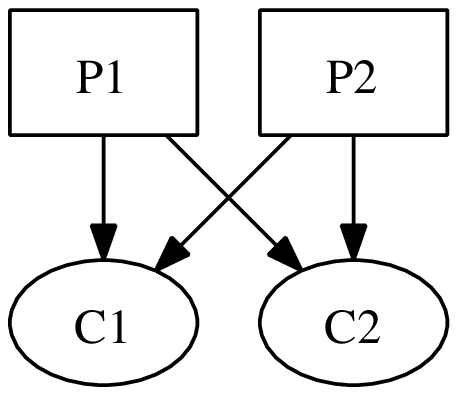}
	\caption{An architecture violating the assertion}
	\label{arch1}
\end{figure}

\begin{figure}[h!t!]
\begin{center}
%\begin{comment}
\setmscscale{0.75}
\begin{msc}{Assertion violation}
\declinst{P1}{}{P1}
\declinst{P2}{}{P2}
\declinst{C1}{}{C1}
\declinst{C2}{}{C2}

\mess{OFFER (1)}{P1}{C1}
\nextlevel
\mess{OFFER (2)}{P1}{C2}
\nextlevel
\mess{OFFER (3)}{P2}{C1}
\msccomment[r]{n=1}{C2}
\nextlevel
\mess{OFFER (4)}{P2}{C2}
\msccomment[r]{n=2}{C2}
\nextlevel
\mess{LOCK (5)}{C1}{P1}
\nextlevel
\mess{OK (6)}{P1}{C1}
\nextlevel
\mess{LOCK (7)}{C1}{P2}
\nextlevel
\mess{OK (8)}{P2}{C1}
\nextlevel
\mess{LOCK (9)}{C2}{P1}
\nextlevel
\mess{START (10)}{C1}{P1}
\nextlevel
\mess{REFUSE (11)}{P1}[0]{C2}[7]
\nextlevel
\mess{START (12)}{C1}{P2}
\nextlevel
\mess{REFUSE (13)}{P2}{C2}
\nextlevel
\mess{ACKREF (14)}{C2}{P2}
\nextlevel
\mess{UNLOCK (15)}{C2}{P1}
\msccomment[r]{n=0}{C2}
\nextlevel
\mess{OFFER (16)}{P2}{C1}
\nextlevel
\mess{OFFER (17)}{P2}{C2}
\msccomment[r]{n=1}{C2}
\nextlevel
\msccomment[r]{n=0}{C2}
\nextlevel
\mess{ACKREF (18)}{C2}{P1}
%\nextlevel
%\mess{OFFER (19)}{P1}{C1}
%\nextlevel
%\mess{OFFER (20)}{P1}{C2}
%\msccomment[r]{n=1}{C2}
%\nextlevel
\end{msc}
\end{center}
\caption{\label{msc} A Message Sequence Chart showing the 
counterexample for the $\alpha$-core protocol}
%\vspace{-0.2cm}
\end{figure}
%\end{comment}

\section{A Tool for Genetic Programming Based on Model Checking}

We constructed a tool, MCGP~\cite{KP6}, that implements our ideas about
model checking based genetic programming.
Depending on these settings, the tool can be used for several purposes:
\begin{itemize}
\item Setting all parts as \emph{static} will cause the tool to just run the
enhanced model
checking algorithm on the user-defined program, and provide its detailed
results.
\item Setting the \emph{init} process as \emph{static} and all or some 
of the other processes as \emph{dynamic} will order the tool 
to synthesize code according to the specified architecture.
This can be used for synthesizing programs from scratch, synthesizing only some
missing parts of a given partial program, or trying to correct or improve a complete given program.
\item Setting the \emph{init} process as \emph{dynamic} and all other processes as static, 
is used when trying to falsify a given parametric program by searching for a configuration
that violates its specification (see~\cite{KP5}).
\item Setting both the \emph{init} and the program processes as \emph{dynamic}
is used for
synthesizing parametric programs, where the tool alternatively evolves various 
programs and configurations under which the programs have to be satisfied.

%\item Setting the \emph{init} process as dynamic can be used for parameterized synthesis, 
%where the tool alternatively generates programs and a set of architectures on which this 
%programs should satisfy the specification (see~\cite{KP5}).

\end{itemize}
\begin{figure}
\begin{center}
\includegraphics[scale=0.60]{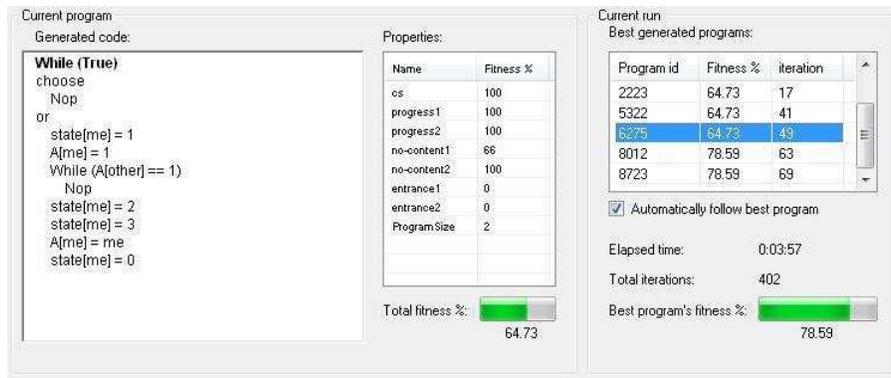}
%\vspace{-1.3cm}
\caption{MCGP screen shot during synthesis of a mutual exclusion algorithm}
\label{screen}
\end{center}
\end{figure}
%\vspace{-0.5cm}

\section{Replacing Model Checking by SAT Solving}

Our approach can use automated deductive techniques instead of
model checking in order to prove the 
correctness of the synthesized algorithms.
However, it requires the verification procedures to be both fully 
automatic, and quite fast, so it can be
repeated a large number of times. Obviously, most theorem provers that require some user interaction during the proof process
cannot be used along with our framework. Furthermore, verification in this case is in general undecidable, 
so fast and complete procedure is not achievable.

Recently, there is a growing use of \emph{SAT} and \emph{SMT solvers} for 
verification purposes.
These tools can function as high performance, and light-weight theorem provers for a broad range of decidable theories over first order logic, such as those of equalities with uninterpreted functions, bit-vectors and arrays.
If we restrict our domain and structure of the synthesized programs as shown later, we can successfully 
(and quite quickly) verify their correctness for all inputs in the related domains. 
For some theories, variables are theoretically unbounded, while for other theories, we must limit their width.

Our work is inspired by \cite{Gul1}, in which a set of short but ingenious and nontrivial programs, 
selected from the book \emph{Hacker's Delight} \cite{hack}, were successfully synthesized.
These programs are loop-free, and use expressions over the decidable theory of bit-vectors. Thus, they
can be easily converted into first order formulas which can then be verified by an SMT solver.
The theory of bit-vectors is decidable only when limiting the width of its related variables. 
From a practical point of view, this does not impose a real
constraint, since we can easily check the correctness of programs even with 128-bit variables.
Unlike \cite{Gul1}, we do not use the SMT solver for the direct synthesis of programs. Instead, we
generate and evolve programs using our GP engine, and integrate the SMT solver into our verification component.

We modify our original framework in order to adopt it to the synthesis of sequential programs. 
In this new framework, the configuration provided by the user to the 
GP engine includes a set of building blocks, 
such as variables and functions that are related to the theory in use.
Only loop-free programs are generated.
The specification provided by the user consists of first 
order logic formulas describing pre and post-conditions over the 
above variables.
A new verification component is built for dealing with 
sequential programs, including two modules.
A \emph{Prover} module is able to get programs from the GP engine, 
and transfer them into logical formulas that are then
checked for correctness by the SMT solver against the specification. 
The results received from the SMT solver are then used for
calculating the fitness function, and for generating counterexamples. 
The core of this module is based on the 
\emph{Microsoft Z3} SMT Solver~\cite{z3}.
A \emph{Runner} module is able to run programs directly, and 
check their correctness for specific given test cases.

%\begin{figure}[!htb]
	%\centering
		%\includegraphics[scale=0.6]{newfw}
	%\caption{A framework for the synthesis of sequential programs}
	%\label{newfw}
%\end{figure}

For sequential programs, we can use the Hoare notation
$\{ \varphi \} P \{ \psi \}$ to denote the requirement that if the
execution of the program starts with a state satisfying the
(first order formula) $\varphi$, upon termination, it satisfies $\psi$.
The formula $\varphi$ is over the input variables. Assume they 
do not change. Otherwise, we can use an additional copy of them; 
a fixed part of the code copies them to the changeable copy. The formula
$\psi$ represents the connection between the input and output variables
upon termination.
Termination is not an issue here, as our generated loop-free programs
must always terminate by construction (they contain no loops).
Let $\varphi$ be the common precondition and $\psi_{1}..\psi_n$ be a set 
of post-conditions.
We want to check for each $\psi_i$ whether $\{\varphi\} P \{\psi_i\}$ holds. 
Using standard construction, we obtain a formula $\eta_P$ that
represents the relationship between the input and output variables.

For each postcondition $\psi_i$ we define
\[F_i := \varphi \land \eta_P \land \lnot \psi_i\]
and
\[F'_i := \varphi \land \eta_P \land \psi_i \]

We can define the following three fitness levels in order of
increasing value:
\begin{enumerate}
	\item {\em $F'_i$ is not satisfiable.}
	Then, the program $P$ is incorrect (w.r.t. $\psi_i$) 
	for all possible inputs.
	\item {\em both $F_i$ and $F'_i$ are satisfiable.}
	There exists an input for which $P$ satisfies $\psi_i$.
	\item {\em $F_i$ is unsatisfiable.}
	$P$ is correct (for all inputs).
\end{enumerate}}

As an example for using our basic method, we tried first to synthesize 
one of the simplest programs from \cite{hack}, which is required 
to output $0$ in the variable $R$ if and only if its input $X$ equals $2^{n}-1$ for some non-negative $n$. The GP engine was allowed
to generate straight line programs, using only \emph{assignment} instructions, bit-vectors related operators 
(such as \emph{and}, \emph{or} and \emph{xor}) constants (\emph{0} and \emph{1}), and variables. Within a few seconds, the following correct program was generated.

%\small
\begin{center}
\begin{minipage}{11.4cm}
\begin{verbatim}
T = X + 1
R = T and X
\end{verbatim}
\end{minipage}
\end{center}
%\begin{center}
%\setlength{\fboxsep}{10pt}
%\fbox{\theverbbox}
%\end{center}
%%\begin{figure}[!htb]
%%	\centering
%%  \setlength{\fboxsep}{13pt}
%	\caption{A program computing the X bitwise or Y}
%	\label{bwor}
%%\end{figure}

%\normalsize

Solutions found by the method described above are guaranteed to be correct 
for every possible input (in the domain
of the variables, such as bit-vectors with a specific width). 
This is a major advantage over solutions generated by traditional GP, 
which can usually guarantee
correctness only for the set of test cases. 
However, using test cases can help in building a smoother fitness function
that can direct the generated programs into gradual improvements. 
Hence, we used for calculating
the fitness function, in addition to 
the above satisfiability based levels, a collection of test cases
that failed on previous selected candidates. Each test case is obtained using
the SAT solver when checking satisfiability of 
$F_i$ on a previous candidate program.
It consists of initial values from which we can run 
new checked candidates. Note that running the code on test cases, using
the runner module, is
faster than applying SAT solving using prover module.

After adding the ability to use test cases, we tried to synthesize a more advanced 
program that is required to compute the floor of the
average of its inputs $X$ and $Y$ without overflowing (which may be caused by simply summing the inputs before dividing by two).
Figure~\ref{avg} shows some of the programs generated during the synthesis process (the logical shift right operator is denoted 
by ``\verb->>-'').

\begin{figure}[!htb]
	\begin{center}
	\begin{minipage}{13cm}
\begin{verbatim}

    R = X + Y           T = X >> 1          R = X and Y
    R = R >> 1          R = Y >> 1          T = X xor Y
                        R = T + R           T = T >> 1
                                            R = R + T

       (a)                 (b)                 (c) 
\end{verbatim}
\end{minipage}
\end{center}
	\caption{Synthesized Programs for Computing \emph{avg(X,Y)}}
	\label{avg}
\end{figure}

Program \emph{(a)} is the naive way for computing the average. However, the addition may cause an overflow, and
indeed the program was refuted by the SMT solver, yielding a counterexample with big inputs.
At the next iteration, program \emph{(b)} was generated. While not overflowing, the program is still incorrect
if both of its inputs are odd, which was reflected by a second counterexample. Finally, the more ingenious 
program \emph{(c)} was generated, and verified to be a correct solution (identical to the one presented in \cite{hack}).

\section{Conclusions}

We suggested the use of a methodology and a tool that perform 
synthesis of programs based on genetic
programming guided by model checking.
Code mutation is at the kernel of genetic programming (crossover is
also extensively used, but we did not implement it).
Our method can be used 
for 
\begin{itemize}
\item synthesizing correct-by-design programs,
\item finding errors in protocols with complicated architectures,
\item automatically correcting erroneous code with respect to a given
specification, and
\item improving code, e.g., to perform more efficiently.
\end{itemize}

We demonstrated our method on the classical mutual exclusion
problem, and were able to find existing solutions, as well as
new solutions.

In general, the verification of parametric systems
is undecidable, and in the few methods that
promise termination, quite severe restrictions are required.
The same apply to code synthesis. Nevertheless, we provide a co-evolution method
for synthesizing parametric systems based on accumulating cases to be checked.
Parameters or architectures on which the synthesis failed before, or
test cases based on previous counterexamples are accumulated to be checked
later with new candidate solutions. As the model checking
itself is undecidable, we finish if we obtain a strong enough evidence
that the solution is correct on the accumulated cases.

We allowed constructing the architecture (processes and the channels
between them) as part of the code that can be mutated. Then the
genetic mutation operation can be used in finding architectures in
which given algorithms fail. This can be used to model check code
with varying architecture, and furthermore, to correct it.

We started recently to look at replacing model checking by SAT and SMT
tools. This provides an efficient alternative for some synthesis problem.
In particular, SMT solvers may succeed in some parametric cases where model
checking fails. 

Although our method does not guarantee termination, neither
for finding the error, nor for finding a correct version of the algorithm, 
it is quite general
and can be fine tuned through provided heuristics in a convenient human-assisted
process of code correction.

%An important strength of the work that is presented here 
%is that it was implemented
%and applied on a complicated published protocol to find and correct an actual
%error. 

%\bibliographystyle{splncs2}
\bibliographystyle{eptcs}
\bibliography{infinity}

\begin{thebibliography}{10}
\providecommand{\bibitemdeclare}[2]{}
\providecommand{\surnamestart}{}
\providecommand{\surnameend}{}
\providecommand{\urlprefix}{Available at }
\providecommand{\url}[1]{\texttt{#1}}
\providecommand{\href}[2]{\texttt{#2}}
\providecommand{\urlalt}[2]{\href{#1}{#2}}
\providecommand{\doi}[1]{doi:\urlalt{http://dx.doi.org/#1}{#1}}
\providecommand{\bibinfo}[2]{#2}

\bibitemdeclare{article}{AK}
\bibitem{AK}
\bibinfo{author}{Krzysztof~R. \surnamestart Apt\surnameend} \&
  \bibinfo{author}{Dexter \surnamestart Kozen\surnameend}
  (\bibinfo{year}{1986}): \emph{\bibinfo{title}{Limits for Automatic
  Verification of Finite-State Concurrent Systems}}.
\newblock {\sl \bibinfo{journal}{Inf. Process. Lett.}}
  \bibinfo{volume}{22}(\bibinfo{number}{6}), pp. \bibinfo{pages}{307--309},
  \doi{10.1016/0020-0190(86)90071-2}.

\bibitemdeclare{book}{gpintro}
\bibitem{gpintro}
\bibinfo{author}{W.~\surnamestart Banzhaf\surnameend},
  \bibinfo{author}{P.~\surnamestart Nordin\surnameend}, \bibinfo{author}{R.~E.
  \surnamestart Keller\surnameend} \& \bibinfo{author}{F.~D. \surnamestart
  Francone\surnameend} (\bibinfo{year}{2001}): \emph{\bibinfo{title}{Genetic
  Programming -- An Introduction; On the Automatic Evolution of Computer
  Programs and its Applications (3rd edition)}}.
\newblock \bibinfo{publisher}{Morgan Kaufmann, dpunkt.verlag}.

\bibitemdeclare{inproceedings}{BT}
\bibitem{BT}
\bibinfo{author}{Yoah \surnamestart Bar-David\surnameend} \&
  \bibinfo{author}{Gadi \surnamestart Taubenfeld\surnameend}
  (\bibinfo{year}{2003}): \emph{\bibinfo{title}{Automatic discovery of mutual
  exclusion algorithms}}.
\newblock In: {\sl \bibinfo{booktitle}{PODC}}, p. \bibinfo{pages}{305},
  \doi{10.1145/872035.872080}.

\bibitemdeclare{article}{one-bit}
\bibitem{one-bit}
\bibinfo{author}{James~E. \surnamestart Burns\surnameend} \&
  \bibinfo{author}{Nancy~A. \surnamestart Lynch\surnameend}
  (\bibinfo{year}{1993}): \emph{\bibinfo{title}{Bounds on Shared Memory for
  Mutual Exclusion}}.
\newblock {\sl \bibinfo{journal}{Information and Computation}}
  \bibinfo{volume}{107}(\bibinfo{number}{2}), pp. \bibinfo{pages}{171--184},
  \doi{10.1006/inco.1993.1065}.

\bibitemdeclare{article}{Dijkstra}
\bibitem{Dijkstra}
\bibinfo{author}{Edsger~W. \surnamestart Dijkstra\surnameend}
  (\bibinfo{year}{1965}): \emph{\bibinfo{title}{Solution of a problem in
  concurrent programming control}}.
\newblock {\sl \bibinfo{journal}{Commun. ACM}}
  \bibinfo{volume}{8}(\bibinfo{number}{9}), p. \bibinfo{pages}{569},
  \doi{10.1145/365559.365617}.

\bibitemdeclare{inproceedings}{EN}
\bibitem{EN}
\bibinfo{author}{E.~Allen \surnamestart Emerson\surnameend} \&
  \bibinfo{author}{Kedar~S. \surnamestart Namjoshi\surnameend}
  (\bibinfo{year}{1995}): \emph{\bibinfo{title}{Reasoning about Rings}}.
\newblock In: {\sl \bibinfo{booktitle}{POPL}}, pp. \bibinfo{pages}{85--94},
  \doi{10.1145/199448.199468}.

\bibitemdeclare{inproceedings}{Gul1}
\bibitem{Gul1}
\bibinfo{author}{Sumit \surnamestart Gulwani\surnameend},
  \bibinfo{author}{Susmit \surnamestart Jha\surnameend},
  \bibinfo{author}{Ashish \surnamestart Tiwari\surnameend} \&
  \bibinfo{author}{Ramarathnam \surnamestart Venkatesan\surnameend}
  (\bibinfo{year}{2011}): \emph{\bibinfo{title}{Synthesis of loop-free
  programs}}.
\newblock In: {\sl \bibinfo{booktitle}{PLDI}}, pp. \bibinfo{pages}{62--73},
  \doi{10.1145/1993498.1993506}.

\bibitemdeclare{inproceedings}{KP1}
\bibitem{KP1}
\bibinfo{author}{Gal \surnamestart Katz\surnameend} \& \bibinfo{author}{Doron
  \surnamestart Peled\surnameend} (\bibinfo{year}{2008}):
  \emph{\bibinfo{title}{Genetic Programming and Model Checking: Synthesizing
  New Mutual Exclusion Algorithms}}.
\newblock In: {\sl \bibinfo{booktitle}{ATVA}}, {\sl \bibinfo{series}{LNCS}}
  \bibinfo{volume}{5311}, pp. \bibinfo{pages}{33--47},
  \doi{10.1007/978-3-540-88387-6\_5}.

\bibitemdeclare{inproceedings}{KP2}
\bibitem{KP2}
\bibinfo{author}{Gal \surnamestart Katz\surnameend} \& \bibinfo{author}{Doron
  \surnamestart Peled\surnameend} (\bibinfo{year}{2008}):
  \emph{\bibinfo{title}{Model Checking-Based Genetic Programming with an
  Application to Mutual Exclusion}}.
\newblock In: {\sl \bibinfo{booktitle}{TACAS}}, {\sl \bibinfo{series}{LNCS}}
  \bibinfo{volume}{4963}, pp. \bibinfo{pages}{141--156},
  \doi{10.1007/978-3-540-78800-3\_11}.

\bibitemdeclare{inproceedings}{KP4}
\bibitem{KP4}
\bibinfo{author}{Gal \surnamestart Katz\surnameend} \& \bibinfo{author}{Doron
  \surnamestart Peled\surnameend} (\bibinfo{year}{2009}):
  \emph{\bibinfo{title}{Synthesizing Solutions to the Leader Election Problem
  using Model Checking and Genetic Programming}}.
\newblock In: {\sl \bibinfo{booktitle}{HVC}}, {\sl \bibinfo{series}{LNCS}}
  \bibinfo{volume}{6405}, pp. \bibinfo{pages}{117--132},
  \doi{10.1007/978-3-642-19237-1\_13}.

\bibitemdeclare{inproceedings}{KP5}
\bibitem{KP5}
\bibinfo{author}{Gal \surnamestart Katz\surnameend} \& \bibinfo{author}{Doron
  \surnamestart Peled\surnameend} (\bibinfo{year}{2010}):
  \emph{\bibinfo{title}{Code Mutation in Verification and Automatic Code
  Correction}}.
\newblock In: {\sl \bibinfo{booktitle}{TACAS}}, \bibinfo{series}{LNCS}, pp.
  \bibinfo{pages}{435--450}, \doi{10.1007/978-3-642-12002-2\_36}.

\bibitemdeclare{inproceedings}{KP6}
\bibitem{KP6}
\bibinfo{author}{Gal \surnamestart Katz\surnameend} \& \bibinfo{author}{Doron
  \surnamestart Peled\surnameend} (\bibinfo{year}{2010}):
  \emph{\bibinfo{title}{MCGP: A Software Synthesis Tool Based on Model Checking
  and Genetic Programming}}.
\newblock In: {\sl \bibinfo{booktitle}{ATVA}}, pp. \bibinfo{pages}{359--364},
  \doi{10.1007/978-3-642-15643-4\_28}.

\bibitemdeclare{article}{Kessels}
\bibitem{Kessels}
\bibinfo{author}{Joep L.~W. \surnamestart Kessels\surnameend}
  (\bibinfo{year}{1982}): \emph{\bibinfo{title}{Arbitration Without Common
  Modifiable Variables}}.
\newblock {\sl \bibinfo{journal}{Acta Inf.}} \bibinfo{volume}{17}, pp.
  \bibinfo{pages}{135--141}, \doi{10.1007/BF00288966}.

\bibitemdeclare{inproceedings}{z3}
\bibitem{z3}
\bibinfo{author}{Leonardo~Mendon\c \surnamestart {c}a~de Moura\surnameend} \&
  \bibinfo{author}{Nikolaj \surnamestart Bj{\o }rner\surnameend}
  (\bibinfo{year}{2008}): \emph{\bibinfo{title}{Z3: An Efficient SMT Solver}}.
\newblock In: {\sl \bibinfo{booktitle}{TACAS}}, pp. \bibinfo{pages}{337--340},
  \doi{10.1007/978-3-540-78800-3\_24}.

\bibitemdeclare{inproceedings}{NPP}
\bibitem{NPP}
\bibinfo{author}{Peter \surnamestart Niebert\surnameend},
  \bibinfo{author}{Doron \surnamestart Peled\surnameend} \&
  \bibinfo{author}{Amir \surnamestart Pnueli\surnameend}
  (\bibinfo{year}{2008}): \emph{\bibinfo{title}{Discriminative Model
  Checking}}.
\newblock In: {\sl \bibinfo{booktitle}{CAV}}, {\sl \bibinfo{series}{LNCS}}
  \bibinfo{volume}{5123}, \bibinfo{publisher}{Springer}, pp.
  \bibinfo{pages}{504--516}, \doi{10.1007/978-3-540-70545-1\_48}.

\bibitemdeclare{book}{SRM}
\bibitem{SRM}
\bibinfo{author}{Doron \surnamestart Peled\surnameend} (\bibinfo{year}{2001}):
  \emph{\bibinfo{title}{Software Reliability Methods}}.
\newblock \bibinfo{publisher}{Springer}, \doi{10.1007/978-1-4757-3540-6}.

\bibitemdeclare{article}{alpha}
\bibitem{alpha}
\bibinfo{author}{Jose~Antonio \surnamestart Perez\surnameend},
  \bibinfo{author}{Rafael \surnamestart Corchuelo\surnameend} \&
  \bibinfo{author}{Miguel \surnamestart Toro\surnameend}
  (\bibinfo{year}{2004}): \emph{\bibinfo{title}{An order-based algorithm for
  implementing multiparty synchronization}}.
\newblock {\sl \bibinfo{journal}{Concurrency - Practice and Experience}}
  \bibinfo{volume}{16}(\bibinfo{number}{12}), pp. \bibinfo{pages}{1173--1206},
  \doi{10.1002/cpe.903}.

\bibitemdeclare{inproceedings}{PetersonF77}
\bibitem{PetersonF77}
\bibinfo{author}{\surnamestart Peterson\surnameend} \&
  \bibinfo{author}{\surnamestart Fischer\surnameend} (\bibinfo{year}{1977}):
  \emph{\bibinfo{title}{Economical Solutions to the Critical Section Problem in
  a Distributed System}}.
\newblock In: {\sl \bibinfo{booktitle}{STOC: ACM Symposium on Theory of
  Computing (STOC)}}, pp. \bibinfo{pages}{91--97}, \doi{10.1145/800105.803398}.

\bibitemdeclare{inproceedings}{PR1}
\bibitem{PR1}
\bibinfo{author}{Amir \surnamestart Pnueli\surnameend} \& \bibinfo{author}{Roni
  \surnamestart Rosner\surnameend} (\bibinfo{year}{1989}):
  \emph{\bibinfo{title}{On the Synthesis of a Reactive Module}}.
\newblock In: {\sl \bibinfo{booktitle}{POPL}}, pp. \bibinfo{pages}{179--190},
  \doi{10.1145/75277.75293}.

\bibitemdeclare{inproceedings}{PR2}
\bibitem{PR2}
\bibinfo{author}{Amir \surnamestart Pnueli\surnameend} \& \bibinfo{author}{Roni
  \surnamestart Rosner\surnameend} (\bibinfo{year}{1990}):
  \emph{\bibinfo{title}{Distributed Reactive Systems Are Hard to Synthesize}}.
\newblock In: {\sl \bibinfo{booktitle}{FOCS}}, pp. \bibinfo{pages}{746--757},
  \doi{10.1109/FSCS.1990.89597}.

\bibitemdeclare{inproceedings}{Tsay}
\bibitem{Tsay}
\bibinfo{author}{Yih-Kuen \surnamestart Tsay\surnameend}
  (\bibinfo{year}{1998}): \emph{\bibinfo{title}{Deriving a Scalable Algorithm
  for Mutual Exclusion}}.
\newblock In: {\sl \bibinfo{booktitle}{DISC}}, pp. \bibinfo{pages}{393--407},
  \doi{10.1007/BFb0056497}.

\bibitemdeclare{book}{hack}
\bibitem{hack}
\bibinfo{author}{Henry~S. \surnamestart Warren\surnameend}
  (\bibinfo{year}{2002}): \emph{\bibinfo{title}{Hacker's Delight}}.
\newblock \bibinfo{publisher}{Addison-Wesley Longman Publishing Co., Inc.},
  \bibinfo{address}{Boston, MA, USA}.

\end{thebibliography}
\end{document}